\documentclass[boxit]{JAC2003}

\usepackage{graphicx}
\usepackage{booktabs}
\usepackage{verbatim}
\usepackage{subcaption}
 {\usepackage[utf8]{inputenc}}           % switch to utf8

\usepackage[USenglish]{babel}
\usepackage{subfloat}

\usepackage{graphicx}
\usepackage{caption}
\usepackage{subcaption}

\begin{document}

\title{Manufacturing and the LLRF Tests of the SANAEM RFQ}

\author{G. Turemen\thanks{gorkem.turemen@ankara.edu.tr}, A. Alacakir, Y. Akgun, S. Bolukdemir, I. Kilic, B. Yasatekin, TAEK-SANAEM, Ankara, Turkey.\\ H. Yildiz, Istanbul University, Istanbul, Turkey.
\\ G. Unel, University of California at Irvine, Irvine, California, USA. }

\maketitle

\begin{abstract}

The Turkish Atomic Energy Authority has been working on building an experimental proton beamline using local resources at the Saraykoy Nuclear Research and Training Center (SANAEM). The radio frequency quadrupole (RFQ) was manufactured  after the completion of beam dynamics and 3D electromagnetic simulation studies. The vanes were machined using a three axis CNC machine. A CMM was used for both the acceptance tests of the vanes and their assembly. Production and assembly errors were found acceptable for this cavity, the very first one developed in Turkey. The aluminum vanes were copper coated by electroplating. The coated vanes were bolted and bonded with eight screws, eight pins and two different adhesives. A silver paste was used for RF sealing and a low vapor pressure epoxy was used for vacuum isolation. First LLRF tests of the RFQ were performed using the phase shift method with a bead-pull setup, a VNA, an N-type RF coupler and a pick-up loop.  Cavity quality factor was measured with 3dB method for different RF sealing stages. This study summarizes the machining, assembling and the first LLRF tests of the SANAEM RFQ.

\end{abstract}

\section{Introduction}
A project to build an experimental proton beamline at Saraykoy Nuclear Research and Training Center of the The Turkish Atomic Energy Authority has been ongoing since 2012. The overview of the design and manufacture of the beamline components has been reported previously\cite{ipac14}. This note focuses on the machining, assembling and the initial low level RF tests of the radio frequency quadrupole, the SANAEM RFQ which was designed using LIDOS software\cite{lidos}. Its beam dynamics and 3D electromagnetic simulation studies were performed using TOUTATIS \cite{toutatis}, DEMIRCI \cite{demirci} and CST \cite{cst}.

\section{Machining of vanes}

\begin{figure}%%[!htb]
   \centering
   \includegraphics*[width=82mm]{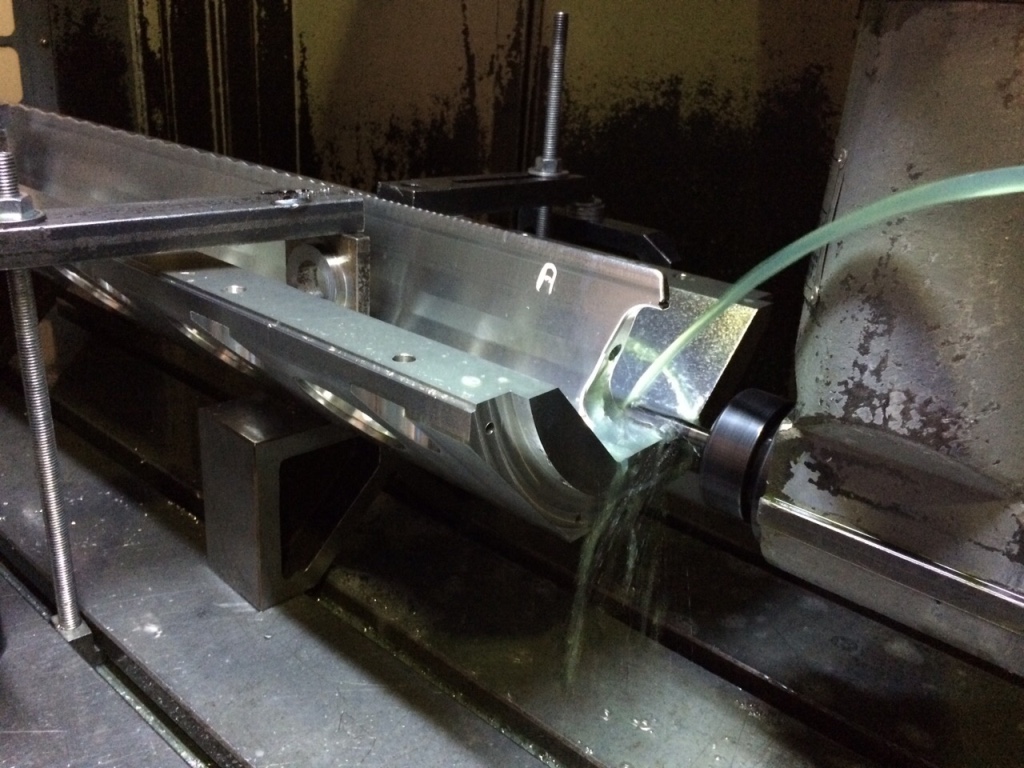}
   \caption{Machining of the cut-backs.}
   \label{mach}
\end{figure}

The production of the vanes was started at early October 2015. After producing a reduced dimension vane and judging the product acceptable, the actual machining of the vanes begun at MAKSAMAK company in Ankara\cite{maksamak}. The vanes were machined from a high-grade aluminum (7075-T6) with a three axis CNC as shown in Fig.\ref{mach}, as it was previously presented in ref \cite{ipac15}. The vanes were produced according to following procedure: Cooling channel drilling, rough machining, port machining, fine machining, cut-back machining, CMM measurements, cooling channel plugging and finishing. The machining process was completed in December 2015.% [hangi tool kullandi sorulacak].  

\subsection{Copper plating}

\begin{figure}%%[!htb]
   \centering
   \includegraphics*[width=82mm]{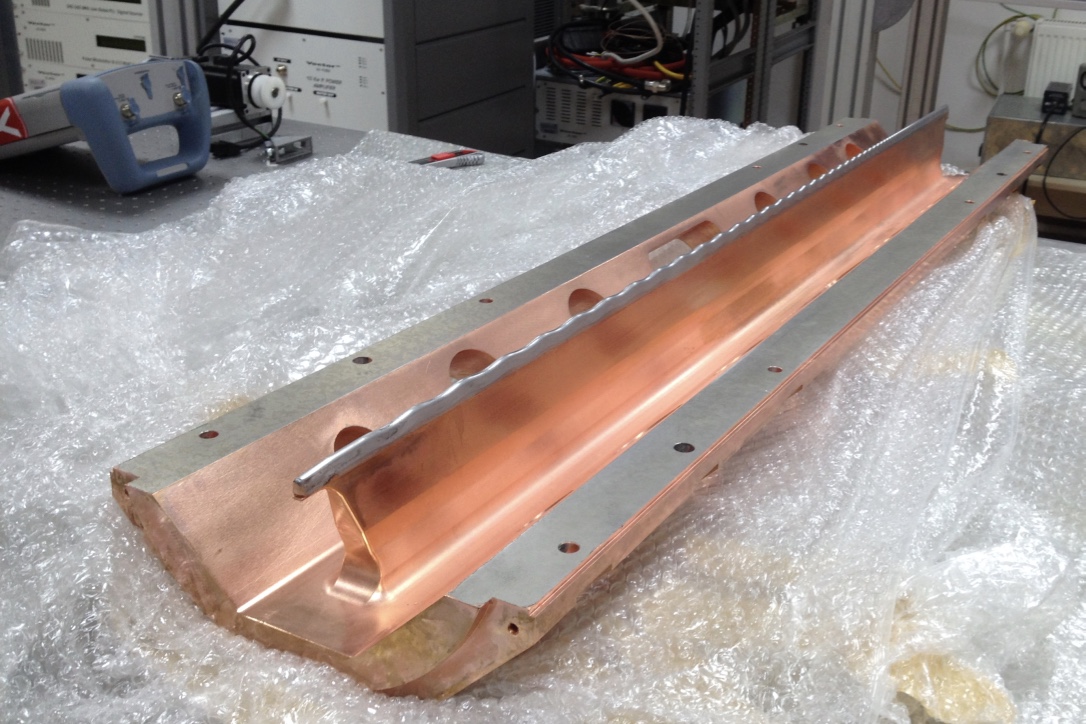}
   \caption{Copper plated vertical vane of the RFQ.}
   \label{plating}
\end{figure}

Following the machining procedure, the vanes were coated with copper using the electroplating method. As the skin depth of copper at 352.21 MHz is about 3.4 $\mu$m, a plating thickness of 30 $\mu$m was considered satisfactory. Therefore all the vanes, tuners and end flanges were plated under the same conditions. The vane-tips were left unplated due to negligible magnetic field amplitude at this region and to preserve their surface smoothness. Similarly, the surfaces of the assembly regions were excluded from the plating process (Fig.\ref{plating}). Additionally, the cooling channels were sealed to restrain corrosion at the operation phase. The plating thickness was observed to be highly variant between the cavity end and center regions due to the RFQ length, 1.2 m.

\subsection{Acceptance tests}
\begin{figure}%%[!htb]
   \centering
   \includegraphics*[width=82mm]{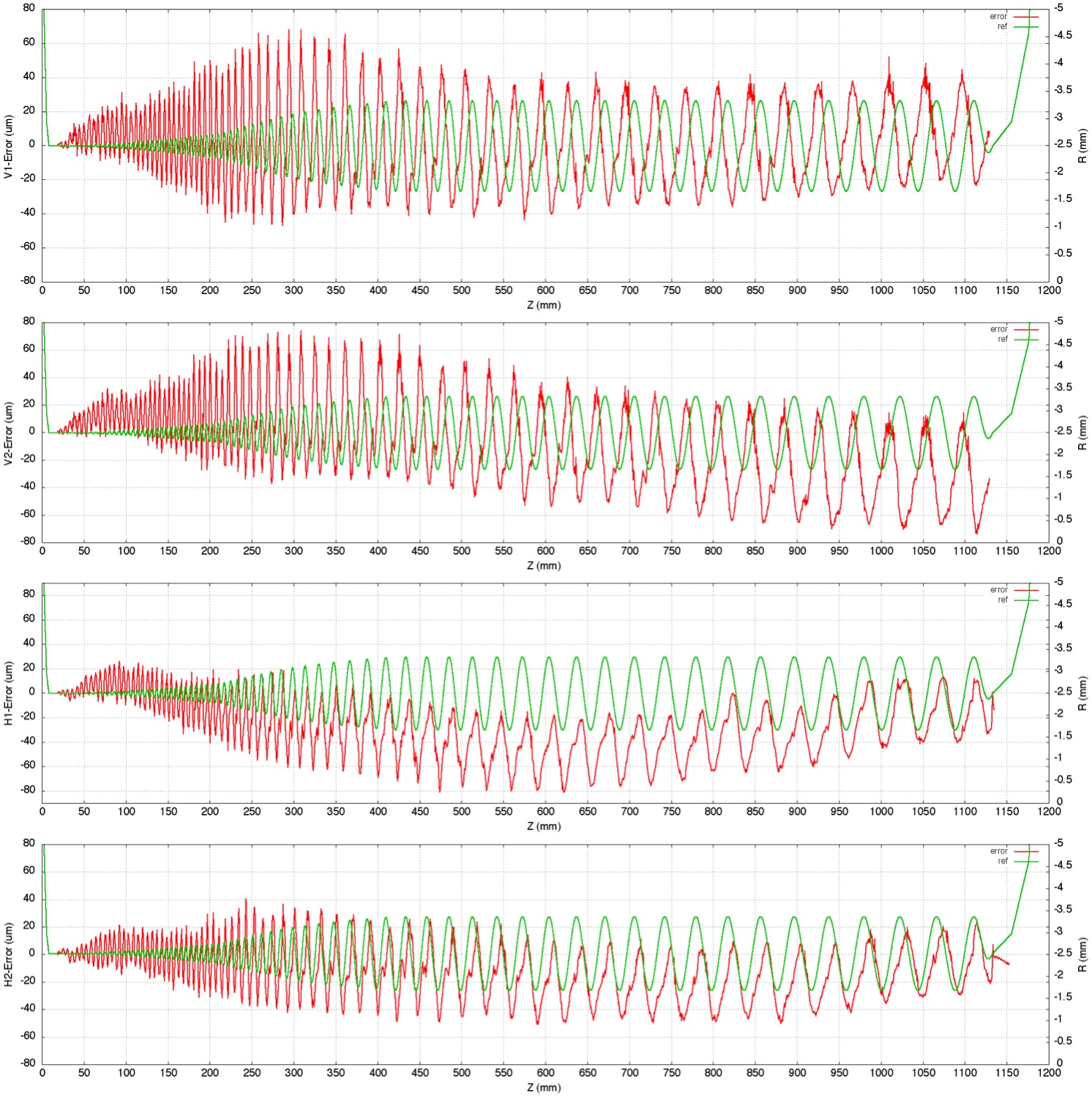}
   \caption{Vane machining errors (in red) and the modulation profiles (in green) for each of the four vanes.}
   \label{err}
\end{figure}

CMM measurements were performed on each vane to verify the machining tolerances. Vane-tip profiles were measured in about 1800 longitudinal steps. The machining errors were found to fluctuate within $\mp$ 80 $\mu$m as can be seen in Fig.\ref{err}.

A number of computer simulations were performed with LIDOS to predict the impacts of these errors on accelerator performance. The results of these simulations are shown in Table \ref{terr}. Therefore the SANAEM RFQ's transmission is expected to be about 78\%.
\begin{table}[!hbt]
   \centering
   \caption{Error analysis.}
   \begin{tabular}{lccccc}
       \toprule
         Aperture error [$\mu$m] & 0 & 20 & 50 & 100 & 200 \\
       \midrule
          Transmission [\%] & 99.2 & 98.3 & 94.4 & 67.2 & 24.6 \\    
       \bottomrule
   \end{tabular}
   \label{terr}
\end{table} 

The CMM test results were found acceptable for a very first RFQ which is developed in Turkey.  The aligning pins on the assembly surfaces of the RFQ were modified according to the error profile, to suppress the banana-like vane shape. After the machining process, the copper regions of the RFQ were cleaned using an alumina suspension with 1 $\mu$m grain size. 

\section{Assembling the vanes}

\begin{figure}%[!htb]
   \centering
   \includegraphics*[width=80mm]{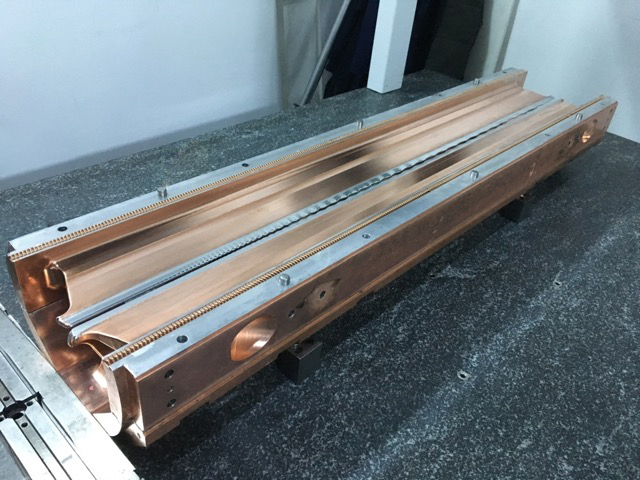}
   \caption{Assembling of the vanes.}
   \label{ass}
\end{figure}

The SANAEM RFQ was machined from aluminum and plated with copper. As  there is no brazing furnace of appropriate length is available locally to assemble the vanes,  a different method consisting of bolting and bonding, was suggested to assemble the vanes \cite{linac14}. Six pin holes were drilled to help the vane alignment in longitudinal and horizontal axes. The vanes surfaces were machined with reference to these pin holes as shown in Fig.\ref{ass}. Then, the copper RF spring seals were placed in their position. Alignment pins were installed to the pin holes and four parts of the RFQ were mounted. The RFQ vanes were pre-assembled with eight brass bolts. CMM measurements were performed to analyze the assembling errors. The errors were compensated using a plastic mallet. All the assembling operations were performed on a CMM table.

\subsection{Acceptance tests}
The aperture errors at the two ends of the RFQ were investigated after the assembling process. The maximum aperture error was found as 44 $\mu$m at the exit-end side of the RFQ. However, in some  locations, the error in the alignment of the reciprocal vanes was found as large as 100 $\mu$m.

\subsection{Bonding, sealing and isolation}
The rigidity of the RFQ was provided by using two different adhesives, after the acceptance tests. Initially, a silver paste was applied to provide a good electrical contact between the vanes for RF sealing. Then, a low vapor pressure epoxy was used to procure the vacuum isolation of the RFQ. 

\section{Tuners, vacuum and RF ports}

There are 16 moveable tuners, 8 vacuum ports and 4 RF ports on the SANAEM RFQ. In spite of the tuners are moveable, they were designed to be vacuum-tight and RF sealed. It is planned to provide the RFQ vacuum with two getter and two turbo pumps. The remaining four of the ports will be empty in the operation of the cavity. All vacuum ports were produced without grills to be used as a tuner in the low level RF (LLRF) tuning. The final bead-pull test is planned to be performed with grills to compensate the additional effects. The RF ports were designed to be modifiable for using a rectangular waveguide coupler or a coaxial loop coupler. 

\section{LLRF tests}

All the LLRF tests of the cavity were performed with vector network analyzer (VNA) \cite{vna}. Two identical loops were used in the tests. The loops were matched to 50 ohm with a matching circuit. Two reciprocal vacuum ports were used to host these loops. The N-type loops were connected to VNA with RG-213 cables.

\subsection{Resonant frequency}

\begin{figure}%[!htb]
   \centering
   \includegraphics*[width=75mm]{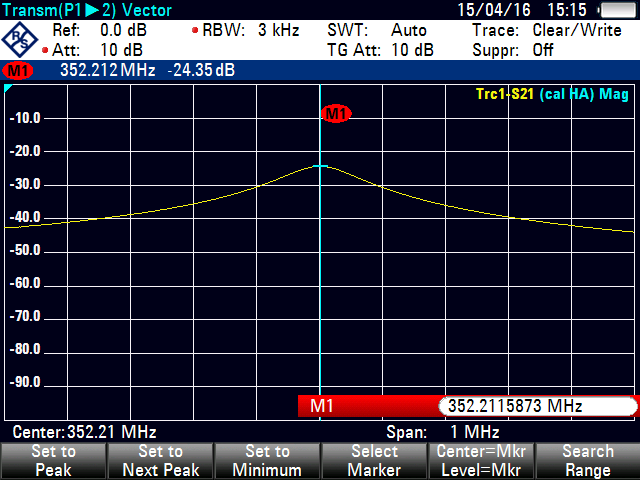}
   \caption{S21 curve from the resonant frequency tuning of the RFQ.}
   \label{tunef}
\end{figure}

The mode spectrum of the cavity was observed from S21 measurements. The resonant frequency of the quadrupole mode was estimated as 351.8 MHz from CST-MWS simulations, while it was measured as 353.4 MHz. The  errors in manufacturing and assembling of the vanes are thought to have led to this frequency shift. However the shift does not cause any problem due the existence of the tuners. RFQ was tuned to the desired frequency of 352.2 MHz by pulling the tuners about 6 mm from the flushed position as shown in Fig.\ref{tunef}.

\subsection{Quality factor}

The intrinsic quality factor ($Q_0$) of the cavity was calculated as 9242 with the CST-MWS simulations. The loaded quality factor ($Q_L$) of the cavity was measured at the two different installation phases, unshielded and shielded. At critical coupling, the $Q_L$ was measured as 2241 and 3213, respectively, without and with RF spring. Therefore $Q_0$ of the RFQ was measured as 6426 which is approximately 70\% of the theoretical value. The copper plating roughness, the  aluminum surface of the eight tuners (which will be plated later) and the vane assembly method are considered to be the causes of this reduction. 

\subsection{Field flatness}

\begin{figure}%[!htb]
   \centering
   \includegraphics*[width=75mm]{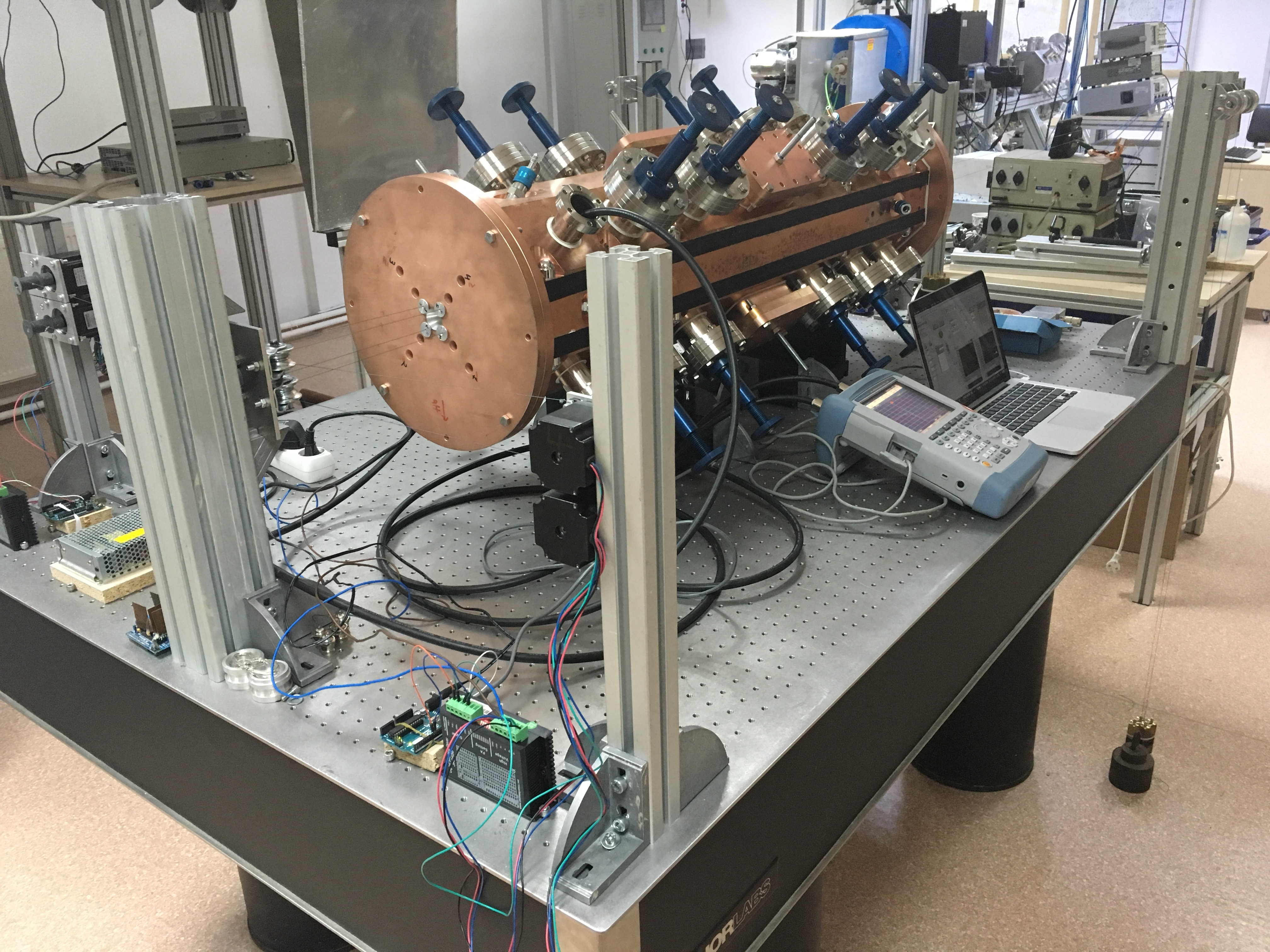}
   \caption{Bead-pull measurement setup.}
   \label{bpull}
\end{figure}

The field flatness of the cavity was calculated with the bead-pull tests on an optical table which can be seen in Fig.\ref{bpull}. Four stepper motors were used to drive the teflon ($\phi$: 10 mm) beads in the cavity with a dyneema wire. The alignment of the beads were provided with 8 hollow cylinders plugged into the cavity flanges. The electric field was measured at 15 mm away from the center.

\begin{figure}%[!htb]
   \centering
   \includegraphics*[width=75mm]{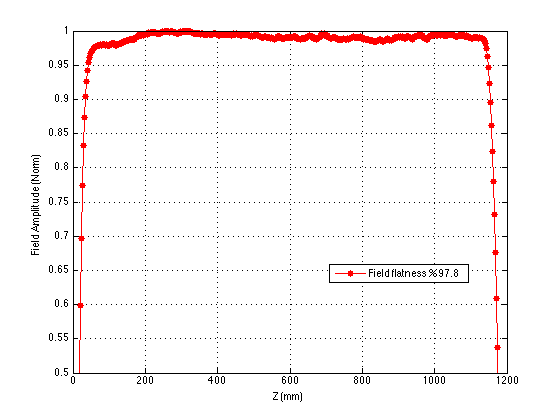}
   \caption{Untuned electric field flatness of the RFQ.}
   \label{ffut}
\end{figure}

The phase shift method was used to calculate the perturbation due to the beads. A LabVIEW \cite{labview} program was written to drive the stepper motors and to analyze VNA data in real-time during the measurements. The untuned field flatness of the cavity was measured as 97.8\% which can be seen from Fig.\ref{ffut}. 

\section{Outlook}
The SPP beamline is slowly taking shape and the first tests with protons are expected for the end of 2016.

\section{Acknowledgments}
This study is funded by TAEK with a project No. A4.H4.P1.

\end{document}